\documentclass[]{spie}  

\usepackage{amsmath,amsfonts,amssymb}
\usepackage{graphicx}
\usepackage[colorlinks=true, allcolors=blue]{hyperref}

\title{SOXS AIT: a paradigm for system engineering of a medium class telescope instrument.}

\author[a]{Riccardo Claudi}
\author[a]{Kalyan Radhakrishnan}
\author[a]{Federico Battaini}
\author[b]{Sergio Campana}
\author[c]{Pietro Schipani}
\author[b]{Matteo Aliverti}
\author[d]{Jos\`e Antonio Araiza--Duran}
\author[a]{Andrea Baruffolo}
\author[e]{Sagi Ben--Ami}
\author[d]{Anna Brucalassi}
\author[c]{Giulio Capasso}
\author[c]{Mirko Colapietro}
\author[f,g]{Rosario Cosentino}
\author[h]{Francesco D'Alessio}
\author[b]{Paolo D'Avanzo}
\author[g]{Rosario Di Benedetto} 
\author[c]{Sergio D'Orsi}
\author[b]{Matteo Genoni}
\author[e]{Ofir Hershko}
\author[i,j]{Hanindyo Kuncarayakti}
\author[b]{Marco Landoni}
\author[g]{Matteo Munari}
\author[k,l]{Giuliano Pignata}
\author[e]{Michael Rappaport}
\author[a]{Davide Ricci}
\author[m]{Adam Rubin}
\author[g,n]{Salvo Scuderi}
\author[o]{Stephen Smartt}
\author[h]{Fabrizio Vitali}
\author[o]{David Young}
\author[g]{Ricardo Zanmar Sanchez}
\author[p]{Jani Achr\'en}
\author[q]{Iair Arcavi}
\author[e]{Rachel Bruch}
\author[a]{Enrico Cappellaro}
\author[c]{Massimo Della Valle}
\author[e]{Avishay Gal--Yam}
\author[r]{Gianluca Li Causi}
\author[a]{Luca Marafatto}
\author[i]{Seppo Matila}
\author[b]{Marco Riva}
\author[a]{Bernardo Salasnich}
\author[s]{Maximilian Stritzinger}

\affil[a]{INAF--Osservatorio Astronomico di Padova, vicolo Osservatorio 5, 35122 Padova, Italy}
\affil[b]{INAF-- Osservatorio Astronomico di Brera, via Bianchi 46, 23807 Merate (LC), Italy}
\affil[c]{INAF--Osservatorio Astronomico di Capodimonte, salita Moiariello 16, 80131 Napoli, Italy}
\affil[d]{INAF--Osservatorio Astronomico di Firenze, Largo E. Fermi 5, 50125, Firenze, Italy}
\affil[e]{Weizmann Institute of Science, Herzl St 234, Rehovot, 7610001, Israel}
\affil[f]{FGG--INAF, TNG, Rambla J.A. Fern\'andez P\'erez 7, 38712 Bre\~na Baja (TF), Spain}
\affil[g]{INAF--Osservatorio Astrofisico di Catania, via di Santa Sofia 78, 95123 Catania, Italy}
\affil[h]{INAF--Osservatorio Astronomico di Roma, via Frascati 33, 00078, Monte Porzio Catone (Roma), Italy}
\affil[i]{Tuorla Observatory, Dep. of Physics and Astronomy, 20014 University of Turku, Finland}
\affil[j]{Finnish Centre for Astronomy with ESO (FINCA), 20014 University of Turku, Finland}
\affil[k]{Universidad Andres Bello, Avda. Republica 252, Santiago, Chile}
\affil[l]{Millennium Institute of Astrophysics (MAS), Nuncio Monsenor Sotero Sanz 100, Providencia, Santiago, Chile}
\affil[m]{ESO, Karl Schwarzschild Strasse 2, D--85748, Garching bei M\"unchen, Germany}
\affil[n]{INAF - Istituto di Astrofisica Spaziale e Fisica Cosmica, Via Corti 12, I-20133 Milano, Italy}
\affil[o]{Astrophysics Research Centre, Queen's University, Belfast, County Antrim, BT7 1NN, UK}
\affil[p]{Incident Angle Oy, Capsiankatu 4 A 29, 20320 Turku, Finland}
\affil[q]{Tel Aviv University, Department of Astrophysics, 69978 Tel Aviv, Israel}
\affil[r]{INAF - Istituto di Astrofisica e Planetologia Spaziali, Via Fosso del Cavaliere, I-00133 Roma}
\affil[s]{Aarhus University, Ny Munkegade 120, D-8000, Denmark}

\authorinfo{Further author information: (Send correspondence to R. Claudi)\\R. Claudi: E-mail: riccardo.claudi@inaf.it}

\begin{document} 
\maketitle

\begin{abstract}
SOXS (SOn of X-Shooter) is a high-efficiency spectrograph with a mean Resolution-Slit product of $\sim 3500$ over the entire band capable of simultaneously observing the complete spectral range 350-2000 nm. It consists of three scientific arms (the UV-VIS Spectrograph, the NIR Spectrograph and the Acquisition Camera) connected by the Common Path system to the NTT, and the Calibration Unit. We present an overview of the flow from the scientific to the technical requirements, and the realization of the sub-systems. Further, we give an overview of the methodologies used for planning and managing the assembly of the sub-systems, their integration and tests before the acceptance of the instrument in Europe (PAE) along with the plan for the integration of SOXS to the NTT. SOXS could be used as an example for the system engineering of an instrument of moderate complexity, with a large geographic spread of the team.
\end{abstract}


\section{INTRODUCTION}
\label{sec:intro}  
The research on transients has expanded significantly in the past two decades, leading to some of the most recognized discoveries in astrophysics (e.g. gravitational wave events, gamma-ray bursts, super-luminous supernovae, accelerating universe). Nevertheless, so far most of the transient discoveries still lack an adequate spectroscopic follow-up. Thus, it is generally acknowledged that with the availability of so many transient imaging surveys in the next future, the scientific bottleneck is the spectroscopic follow-up of transients. Within this context, SOXS aims to significantly contribute bridging this gap. It will be one of the few spectrographs on a dedicated telescope with a significant amount of observing time to characterize astrophysical transients. It is based on the concept of X-Shooter \cite{vernetetal2011} at the VLT but, unlike its “father”, the SOXS science case is heavily focused on transient events. Foremost, it will contribute to the classifications of transients, i.e. supernovae, electromagnetic counterparts of gravitational wave events, neutrino events, tidal disruptions of stars in the gravitational field of supermassive black holes, gamma-ray bursts and fast radio bursts, X-ray binaries and novae, magnetars, but also asteroids and comets, activity in young stellar objects, and blazars and AGN.

\begin{figure} [ht]
   \begin{center}
   \begin{tabular}{c} 
   \includegraphics[height=10cm]{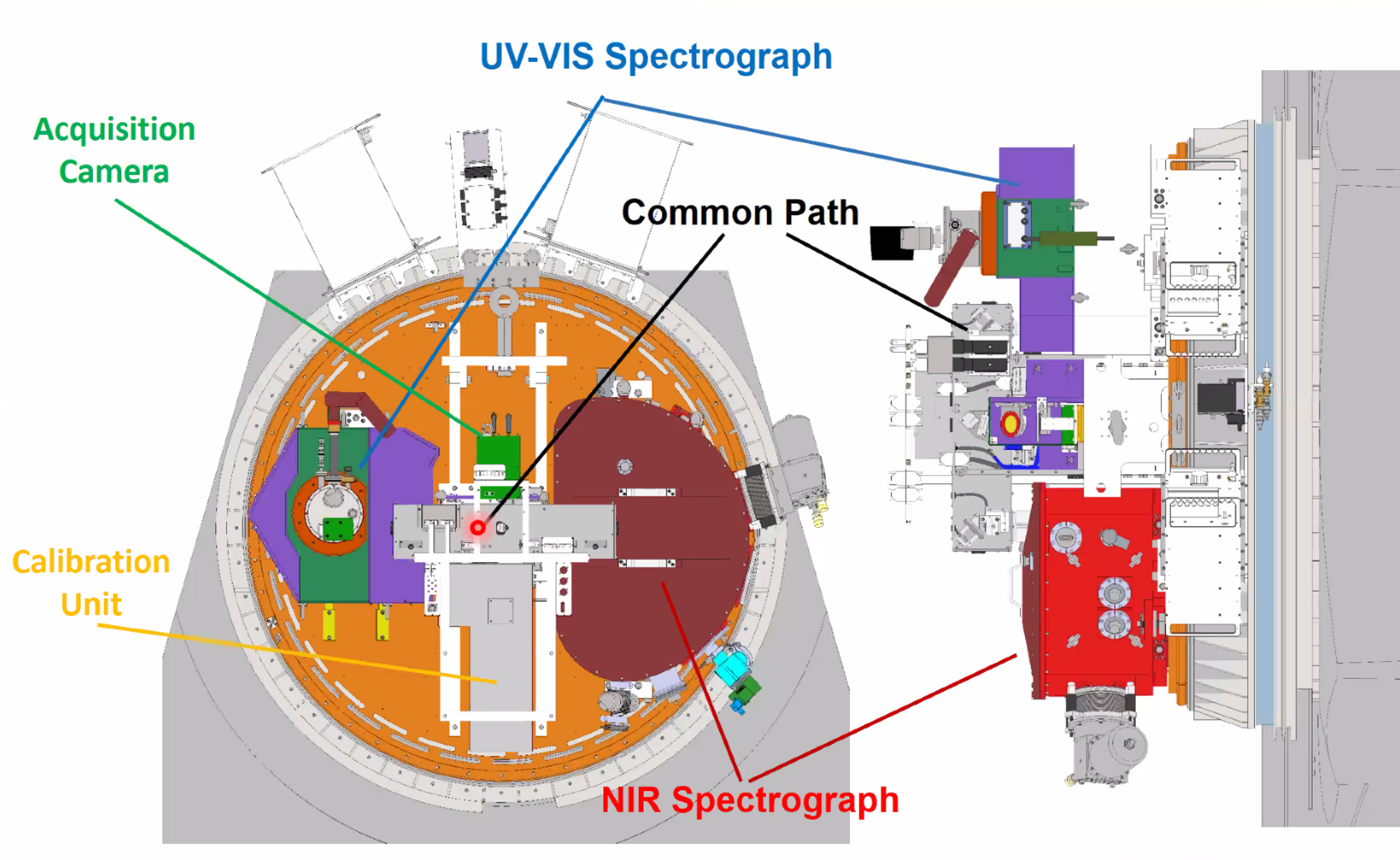}
   \end{tabular}
   \end{center}
   \caption[example] 
   { \label{fig:soxs1} 
SOXS: front and side view of the instrument with the identification of all the sub-systems.}
   \end{figure}

SOXS\cite{schipanietal2018spie, schipanietal2016spie, claudietal2018frap, schipanietal2020spie} will simultaneously cover the electromagnetic spectrum from 0.35 to 2.0\ $\mu$m using two arms (UV--VIS and NIR) with a product slit--resolution of $\sim 4500$. The throughput should enable to reach a S/N$\sim 10$ in a 1-hour exposure of an R=20 mag point source. SOXS, that will see its first light at the end of 2021, will be mounted at the Nasmyth focus of NTT replacing SOFI. The whole system (see Figure\ \ref{fig:soxs1}) is constituted by the three main scientific arms: the UV--VIS spectrograph\cite{rubinetal2018spie,rubinetal2020spie,  cosentinoetal2018spie,cosentinoetal2020spie}, the NIR Spectrograph\cite{vitalietal2018spie, vitalietal2020spie} and the acquisition camera (AC)\cite{brucalassietal2018spie, brucalassietal2020spie}. The three main arms, the calibration box\cite{kuncarayaktietal2020spie} and the NTT are connected together by the Common Path (CP)\cite{claudietal2018spie, biondietal2018spie, biondietal2020spie}. 

The main characteristics of the three scientific arms are listed in Table\ \ref{tab:subsy}.

\begin{table}[ht]
\caption{Main characteristics of the SOXS sub systems connected to the SOXS common path.} 
\label{tab:subsy}
\begin{center}       
\begin{tabular}{|l|c|c|c|} 
\hline
\rule[-1ex]{0pt}{3.5ex}        & \textbf{AG Camera} & \textbf{UV--VIS}& \textbf{NIR}  \\
\hline
\rule[-1ex]{0pt}{3.5ex}  F/\# & 3.6 & 6.5 & 6.5   \\
\hline
\rule[-1ex]{0pt}{3.5ex}  Spectral Range& ugrizY $+$ V& 350--850 nm& 800 -- 2000 nm  \\
\hline
\rule[-1ex]{0pt}{3.5ex}  Resolution& $-$ & $>3600\ \sim4500$\ Avg & 5000  \\
\hline
\rule[-1ex]{0pt}{3.5ex}  Slit Width (arcsec)& $-$ & \multicolumn{2}{|c|}{$0.5 - 1.0 -1.5-5.0$}  \\
\hline
\rule[-1ex]{0pt}{3.5ex}  Slit height (arcsec)& $-$ & \multicolumn{2}{|c|}{$12.0$} \\
\hline
\rule[-1ex]{0pt}{3.5ex}  Pixel Scale (arcsec/px)& 0.205 & 0.280 & 0.164  \\
\hline 
\rule[-1ex]{0pt}{3.5ex}  Detector & Andor Ikon M-934 1k $\times$ 1k& e2v CCD44--82 2k$\times$ 4k& H2RG\ 2k $\times$ 2k  \\
\hline 
\rule[-1ex]{0pt}{3.5ex}  Pixel Size ($\mu$m) & 13.0& 15.0& 18.0  \\
\hline 
\end{tabular}
\end{center}
\end{table}

\section{SOXS AIT management}
\label{sec:aitman}
\begin{figure} [b]
   \begin{center}
   \begin{tabular}{c} 
   \includegraphics[height=8cm]{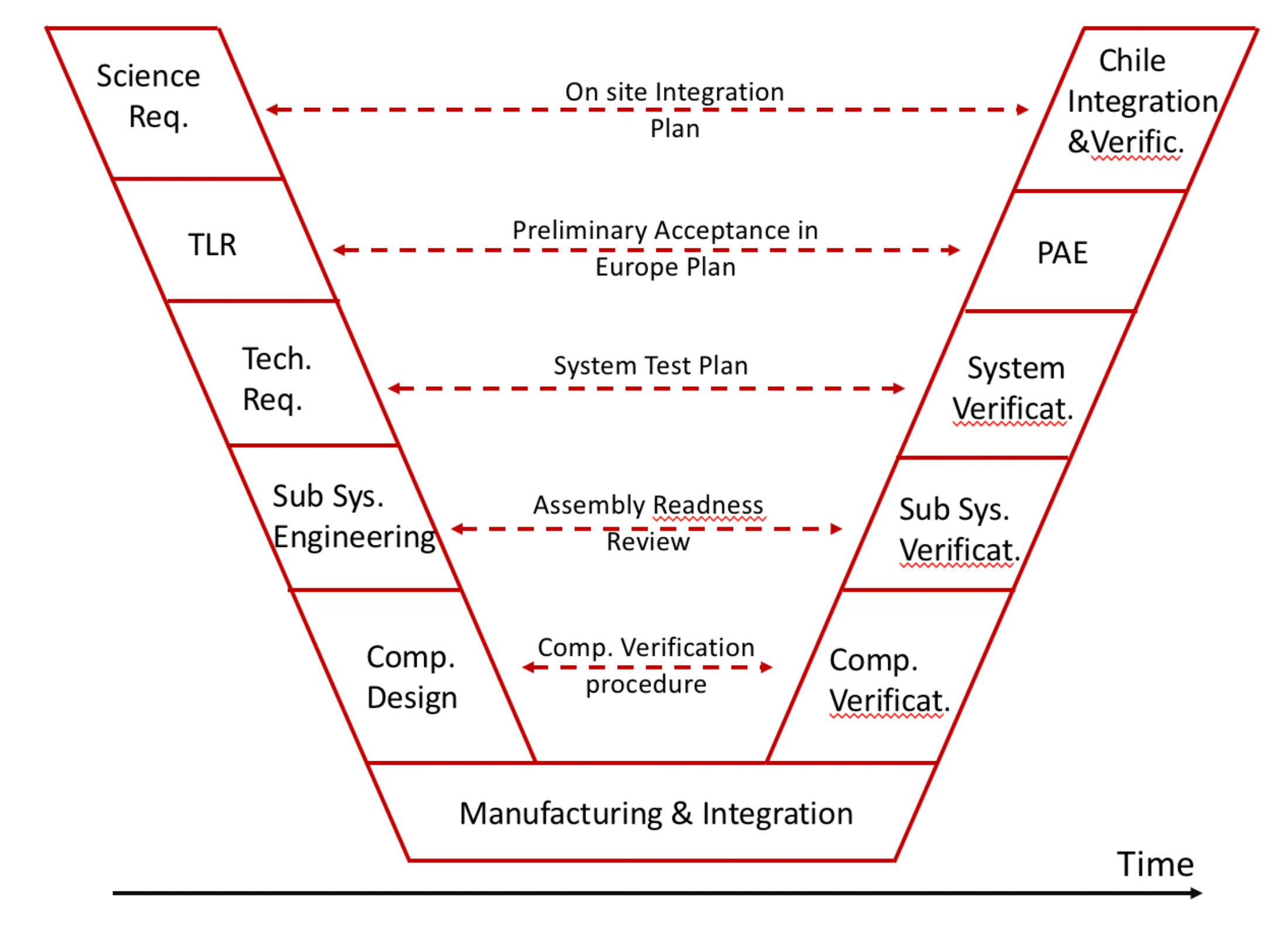}
   \end{tabular}
   \end{center}
   \caption[example] 
   { \label{fig:vdiagram} The typical system engineering V-diagram for SOXS. 
}
   \end{figure}

The instrument construction philosophy is that all sub‐systems will be built in different institutes and will be moved in the integration site before the Preliminary Acceptance in Europe (PAE). Just before the shipping, all sub‐systems have to be tested in order to verify SOXS science and technical requirements (see Section\ \ref{sec:tlr}). This phase is an internal review named Assembly Readiness Review (ARR). After the PAE the instrument will be dismounted and moved to Chile where it will be re‐integrated and tested for acceptance (see Figure\ \ref{fig:vdiagram}).

Eventually, the instrument will undergo to a commissioning phase and science validation phase. During the different phases mainly four type of tests are foreseen:
\begin{description}
\item[Alignment:] are all the procedures that allow having the throughput as high as possible for the sub‐systems and the instrument. It will be necessary to check the alignment every time the sub‐systems and the instrument will be partially dismounted and successively re‐ integrated.
\item[Functional tests:] are all the tests devoted to verify the full functionality of the sub‐systems and of the instrument. These tests will verify the electro‐mechanics movements, sensors and telemetry. The functionality of sub‐systems and the instrument will be tested with both hardware manual controller and INS.
\item[Verification tests:] are all the tests that verify, depending by the test phases, all the internal and external (including those towards the telescope) interfaces. These tests will also verify the compliance of the sub‐systems and the instrument with all environmental requirements described in Section\ \ref{sec:tlr}.
\item[Science tests:] are all the tests that will verify the compliance of the sub‐systems and the instrument with science and those technical requirements that concern the scientific output of the sub‐systems and the instrument.
\end{description}

All these kind of tests will be performed in laboratory, integration hall and at the Nasmyth focus of the telescope. Some tests will be necessary to pass through the different milestones of the SOXS project. Following the flow in Figure\ \ref{fig:vdiagram}, the first set of tests concern the sub systems. Just after the single sub‐system reaches the integration hall (both in INAF – OAPD) and in Chile it will be unpacked and remounted. It will be necessary to check the alignment and confirm the functionality and verification test results reported in the ARR documentation package. Once all the sub‐systems have passed the tests, they will be ready to be integrated on the SOXS flange in the integration hall. 

Once SOXS is integrated in the laboratory (INAF‐OAPD), it shall be tested in order to demonstrate that the instrument fulfills the specifications, and the instrument is ready for shipment to the La Silla observatory.
The Consortium shall demonstrate to ESO the conformity of the instrument performance with the technical requirements and specifications including safety assessment. Tests of electrical safety and electromagnetic compatibility (EMC) will be performed by ESO experts. Based on the results of all tests performed, the Consortium shall prepare a detailed PAE Test Report (PAE). This document is part of the Preliminary Acceptance Data Package and is the basis on which PAE is performed. The data package may be reviewed by a board that might include La Silla staff. If passed, ESO will grant Preliminary Acceptance (Europe) and authorize shipment of SOXS to La Silla. The final PAE meeting should be held at the place where the instrument is located at the contractor premises.

\section{From scientific to technical requirements}
\label{sec:tlr}
The science requirements follow from the science case for SOXS \cite{claudietal2019frapconfe78, schipanietal2018spie} and are the following:
\begin{description}
\item[SR01]: Wavelength range – The wavelength range that is covered by the SOXS in a single shot shall cover at least from u to H bands (350-1750 nm, 320-2050 nm goal).
\item[SR02]: Detector Quantum Efficiency – The instrument shall be highly efficient to be able to obtain quantitative spectroscopic information on the faintest, point-like targets reachable at a 3.5-m telescope at a resolution where the spectrum is just sky limited in about one hour of exposure time in the regions free of emission lines. The absolute efficiency shall be the one given by state of the art coatings, detectors and transmitting materials.
\item[SR03]: Spectral resolution – It shall be intermediate to maximise sensitivity but still be high enough to allow quantitative work on narrow emission and absorption lines under median La Silla seeing and to be relatively unaffected by atmospheric OH emission lines. Sky background, detector noise and OH line density considerations require that the resolution slit product is $>3500$ to permit accurate sky subtraction.
\item[SR04]: Spectral format – The spectral format is cross-dispersed (or quasi), dictated by the requirements on spectral resolution and wavelength range. Efficiency considerations strongly favour prism cross- dispersion.
\item[SR05]: Slit length – It shall be adequate to perform sky subtraction with sufficient accuracy within the constraints of the spectral format. A minimal length of 10 arcsec is envisaged.
\item[SR06]: ADC – In view of its wide band-pass, the instrument shall be equipped with an ADC (Atmospheric Dispersion Compensator) for at least the UV-VIS spectral ranges for good spectrophotometric accuracy and the flexibility to orient the slit as required by the observing program.
\item[SR07]: Observing efficiency – The instrument shall be designed for rapid acquisition and low observing and calibration overheads. There shall be no need to perform calibration during the night. 
\item[SR08]: Calibration system – SOXS shall be equipped with its own calibration system to perform scientific and technical calibrations needed to remove the instrument signature and to maintain the instrument.
\end{description}
From this set of science requirements descend the technical requirements that guided the design of the SOXS Instrument and of its sub-systems. The technical requirements are numbered with a code (TR\#.\#\#) that contains the work package of reference followed by a number in sequence. In the following, we list the technical requirements of SOXS and the kind of verification required at PAE (preliminary acceptance in Europe), before the instrument will be sent in at La Silla (Chile) for the integration at the NTT.
\subsection{Optical Requirements (01)}
\begin{description}
\item[TR01.01] Efficiency: $> 90$\% in the wavelength range $350 - 1750$ nm. Losses would be due to dichroic crossover that are not recoverable by DRS, and order gaps that may be present in orders that are longer than the width of the detectors. Losses due to sky lines are not included but are estimated to be 10‐ 20\% above 650 nm
\item[TR01.02] Wavelength coverage: DQE$_{\text{av}}$ $> 25$\%, $350 - 1750$\ nm (average of order centers), DQE $> 8$\%, $350 - 1750$ nm (any wavelength within FSR; goal $> 10$\%)
\item[TR01.03] Dichroic: the dichroic will have the crossover wavelength at $800 \pm 12$\ nm with a crossover width 30 nm. No ripple or spikes $< 300$ km/sec.
\item[TR01.04] Limiting Magnitude: Limiting Magnitudes, stellar objects, long slit (1h, S/N=10, slit 1.0”, overall thermal emissivity$=25$\%). V$>20$; J$>19$.
\item[TR01.05] Limiting magnitudes on extended objects, long slit: Not specified.
\item[TR01.06] Acquisition Camera system: 
\begin{itemize}
\item Centroid accuracy: $<0.1$” in ubgrizVY; 
\item FoV $2' \times 2'$ (goal $4' \times 4'$);
\item detector QE: $0.5 @0.38\ \mu$m; $0.82 @0.58\ \mu$m ; $0.5 @0.82\ \mu$m
\end{itemize}
\item[TR01.07] Spectral Resolution: R $> 3500$ (goal 5000)
\item[TR01.08] Slit Dimension and Alignment. for both the spectrograph UV-VIS and NIR there are 4 slit positions: 1) slit $0.5 \times 10$”; 2) slit $1.0 \times 10$”; 3) slit $1.5 \times 10$"; 4) slit $5.0 \times 10$”. With the instrument at zenith, the projected slit positions of the NIR spectrographs shall coincide with the projected position of the UV-VIS slit: to 0.05” perpendicular to slit; to 0.2” along the slit
\item[TR01.09] Inter order separation. The interorder separation is $> 0.6$ arcseconds anywhere in each order, over a range corresponding to $1.2 \times$ the FSR.
\item[TR01.10] Optical Efficiency of Preslit $> 80$\%, it includes ADC; efficiency in the dichroic crossover ranges may be summed.
\item[TR01.11] Stray light and ghosts Requirement for science spectra and flatfields: no focussed ghost $> 10^{‐3}$ of neighbouring continuum. Stray light background shall be smooth and $< 5$\% of neighbouring continuum (target: $< 1$\%). Some ghosting is allowed in wavelength calibration spectra.
\item[TR01.12] Synchronous exposure in the two arms. The central time of exposure (i.e. the time at which half of exposure time passed) in the two arms shall be coincident to better than 1 second.

\end{description}
\subsection{Mechanical Requirements (02)}
\begin{description}
\item [TR02.01]Instrument mass. The maximum possible load condition will be 2000 kg at 270 mm from the fixation flange of the instrument.
\end{description}

\subsection{Electronics and EMC Requirements (03)}
\begin{description}
\item [TR03.01] Electronics. The instrument control electronics shall comply with the general requirement documents referenced in the ESO document titled "Requirements for Scientific instruments on the VLT Unit telescopes (VLT-SPE-ESO-10000-2723, Issue 1 dd. 18.03.05).
\item [TR03.02] EMC: the design guidelines for EMC given in VLT Unit telescopes (VLT-SPE-ESO-10000-2723, Issue 1 dd. 18.03.05) constitute good practice for EMC design.
\item [TR03.03] Monitoring of the instrument. The ICE shall perform monitoring of the status of the instrument, reporting values of parameters to the Instrument Control Software (ICS)
\item [TR03.04] Safety procedures. Adequate reaction to: loss of power; loss of coolant; failure of critical AD1 sensors; “emergencies”.

\end{description}

\subsection{UV-VIS Detector Requirements (04)}
The technical requirements of the CCD detector of the UV-VIS spectrograph are listed in Table\ \ref{tab:ccduv}. They are combined in technical requirement 04 (TR04).
\begin{table}[t]
    \centering
    \begin{tabular}{|cc|cc|}
\hline
\textbf{N. Pixel}& $2048 \times 2048$                        & \textbf{Operating T} & $150 - 160$\ K\\
\textbf{Pixel size} &  15.0  $\mu$m                        & \textbf{Dark Current}& $< 2$e$^{-1}$/Pixel\\
\textbf{QE}                  &$>70$\%@300 nm,$>75$\%@320 nm, & \textbf{Read out Modes}& Slow (50 – 100 kpix/sec, tbc)\\
                    &$>77$\%@350 nm,$>78$\%@370 nm, &   & Fast, (600 kpix, sec tbc) \\
                    &$>80$\%@400 nm,$>85$\%@450 nm, & \textbf{Binning options}&  $1 \times 1$, $1 \times 2$, $2 \times 2$    \\
                    &$>70$\%@500 nm,$>75$\%@550 nm, & \textbf{R.O.N.}  & Slow: ($50 - 100$ kpix/sec) $< 3$\ e \\
                    &$>80$\%@600 nm,$>80$\%@650 nm, &   & Fast (600 kpix/sec) $< 8$\ e\\
                    &$>85$\%@700 nm,$>85$\%@750 nm, & \textbf{Pixel saturation}   & $>120000$\ e\\
                    &$>85$\%@800 nm,$>85$\%@850 nm, & \textbf{Linearity}   & $<1$\% $10 - 10^5$ e\\
                    &$>75$\%@900 nm,$>50$\%@950 nm, & \textbf{Cosmetics}   & 1 bad column\\
                    &$>30$\%@1000 nm                &             &  $<0.01$\% hot or dead pixels\\
\hline
    \end{tabular}
    \caption{TR04: Technical requirements for the CCD detector of the UV-VIS spectrograph.}
    \label{tab:ccduv}
\end{table}

\subsection{NIR Detector Requirements (05)}
The technical requirements of the infrared detector of the NIR spectrograph are listed in Table\ \ref{tab:nirdet}. They are combined in technical requirement 05 (TR05).
\begin{table}[b]
    \centering
    \begin{tabular}{|cc|}
\hline
\textbf{N. Pixel}& $2048 \times 2048$ \\
\textbf{Pixel size} &  18.0  $\mu$m  \\
\textbf{QE}                  &$>70$\% in Z\\
                             &$>70$\% in J   \\
                             &$>80$\% in H \\
\textbf{Operating T} & $< 82$\ K \\
\textbf{Dark Current}& $< 10^{-2} - 2 \times 10^{-1}$ e$^{-1}$/Pixel/s\\
\textbf{R.O.N.}      & $< 15$ e (single readout) \\
                     & $<5$ e (Multiple nondestructive)\\
\textbf{Pixel saturation}   & $>80000$\ e \\
\textbf{Linearity}   & $<1$\% \\
\hline
    \end{tabular}
    \caption{TR05: Technical requirements for the detector of the NIR spectrograph.}
    \label{tab:nirdet}
\end{table}

\subsection{Vacuum and Cryogenic Requirements (06)}
\begin{description}
\item [TR06.01] Evacuation time shall be $< 6$ hr from 1 bar to $10^{‐4}$ bar end pressure $10^{‐5}$ bar
\item [TR06.02] Vacuum pressure (cold) $<3 \times 10^{‐6}$ mbar for 6 months or more.
\item [TR06.03] Cool down time shall be $< 30$ hr to operational condition.
\item [TR06.04] The temperatures inside the NIR spectrograph shall be: 1) NIR detector camera lens $40 \pm2$\ K; 2) NIR optical bench: $150 \pm 2$\ K (gradient $< 5$\ K); 3) Prism assembly: $150 \pm 0.1$\ K.
\end{description}

\subsection{Verification of Technical Requirements}
\label{sec:veri}
The goal of the verification procedures is to provide the necessary control to demonstrate that the performance, interface, and safety specifications are met during the Final Design, the Manufacturing, the Integration, and the Commissioning phases. For each technical requirement a different level of verification is defined as function of the several project phases. So far, SOXS passed through the final design review (FDR) and is going towards the preliminary acceptance in Europe (PAE). The two phases are very different. In the former the instrument is only an idea, while in the latter it is built and integrated into an integration hall. So there are different procedures to verify a technical requirement. Starting from the beginning to the end of project, it is possible to identify at least six different verification procedures:
\begin{enumerate}
\item By design (D), a design procedure is adopted to meet the performance
\item By analysis (A), a design analysis is performed and compared to the design performance requirements
\item By inspection (I), an inspection will be performed to verify compliance with the  specification.
\item By test at Component Level (TC), a test will performed on components to demonstrate compliance with the specification
\item By test at Sub System Level (TS), a test will performed on sub – system (e.g. single spectrograph) to demonstrate compliance with the specification
\item By test on the Instrument (TI), test will performed on the complete instrument to demonstrate compliance with the specification
\end{enumerate}
Table\ \ref{tab:vertab} lists the different procedures for the verification of the TR of SOXS described in Section\ \ref{sec:tlr} at PAE.
\begin{table}[t]
    \centering
    \begin{tabular}{|cc|cc|cc|}
    \hline
    TR   & Ver. &    TR   & Ver.&    TR   & Ver.\\
         & Proc.&         & Proc.&         & Proc.\\
    \hline\hline
    TR01.01&  TS/TI & TR01.09 & TS/TI & TR03.04 & TI\\
    TR01.02& TS/TI  & TR01.10 & TS & TR04   & TC/TI \\ 
    TR01.03 & TC    & TR01.11 & TI & TR05   & TC/TI \\
    TR01.04 & A     & TR01.12 & TI & TR06.01 & TS/TI \\
    TR01.05 & A     & TR02.01 & TS &  TR06.02& TS\\
    TR01.06 & TC    & TR03.01 & TI &  TR06.03 & TI\\
    TR01.07 & TS/TI & TR03.02 & TI & TR06.04 & TI\\
    TR01.08 & TI    & TR03.03 & TI &         &    \\
    \hline
    \end{tabular}
    \caption{The different verification procedure of the technical requirements of SOXS at PAE}
    \label{tab:vertab}
\end{table}

\section{SOXS Integration plan at La Silla}
\begin{figure} [t]
   \begin{center}
   \begin{tabular}{c} 
   \includegraphics[height=8cm]{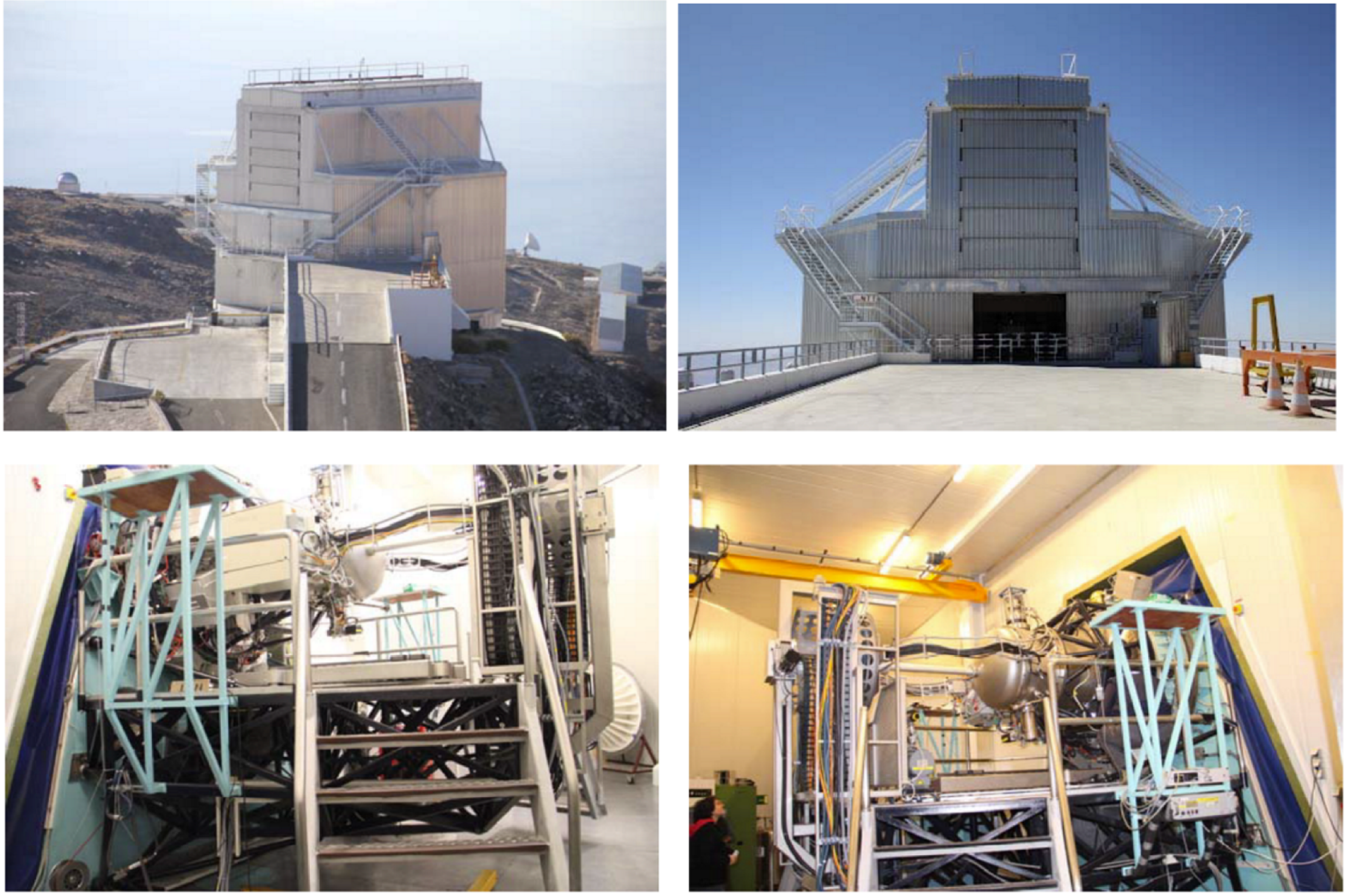}
   \end{tabular}
   \end{center}
   \caption[example] 
   { \label{fig:ntt} The telescope and the Nasmyth-A room @ NTT ESO-La Silla with SOFI, the instrument that will be replaced by SOXS 
}
   \end{figure}

After the PAE, SOXS will be dismount in the inverse order w.r.t. the mounting to the flange. The sub-systems will be packed in their re-usable vibration-isolated crates already used to ship the sub-system in Padova and delivered to the telescope site inside a container (TBC). We will unscrew and send the flange \& column system too. We expect that this operation will last two weeks, three for the shipping. 
SOXS will be hosted in the Nasmyth-A room of the NTT telescope at ESO‐La Silla (Figure\ \ref{fig:ntt}), replacing the SOFI imager/spectrograph. 

The Container coming from Europe will be positioned on the cement balcony (with a capacity of 2 tons and dimensions 1.54 m by 2 m and 2 m height) in front of the ingress door at the telescope. The boxes containing the structure and SOXS sub-parts with no optics inside, will be opened before to enter in the telescope. Instead, sub-system boxes will enter NTT building from the main large door and then a $1.54\ \text{m} \times 2.05$\ m door. Then there is another bottleneck of 1.54 m, before entering the Nasmyth room. In the Nasmyth room a crane and a forklift are available. The temperature of the Nasmyth room is controlled; water-cooling line is available to cool down the electronic racks.

As already pointed out, the integration of SOXS in laboratory conditions allows to learn the peculiar critical issues about the procedure foreseen and to produce or modify the necessary equipment in the cushy environment of our workshop. Moreover, the first integration allows to define the correct shimming of the KM of all the sub-system, so the simply re-coupling of the sub-systems should provide the same I/F flange, column, sub-system configuration reached in Europe, within the repeatability of the KM. 

   \begin{figure} [ht]
   \begin{center}
   \begin{tabular}{c} 
   \includegraphics[height=8cm]{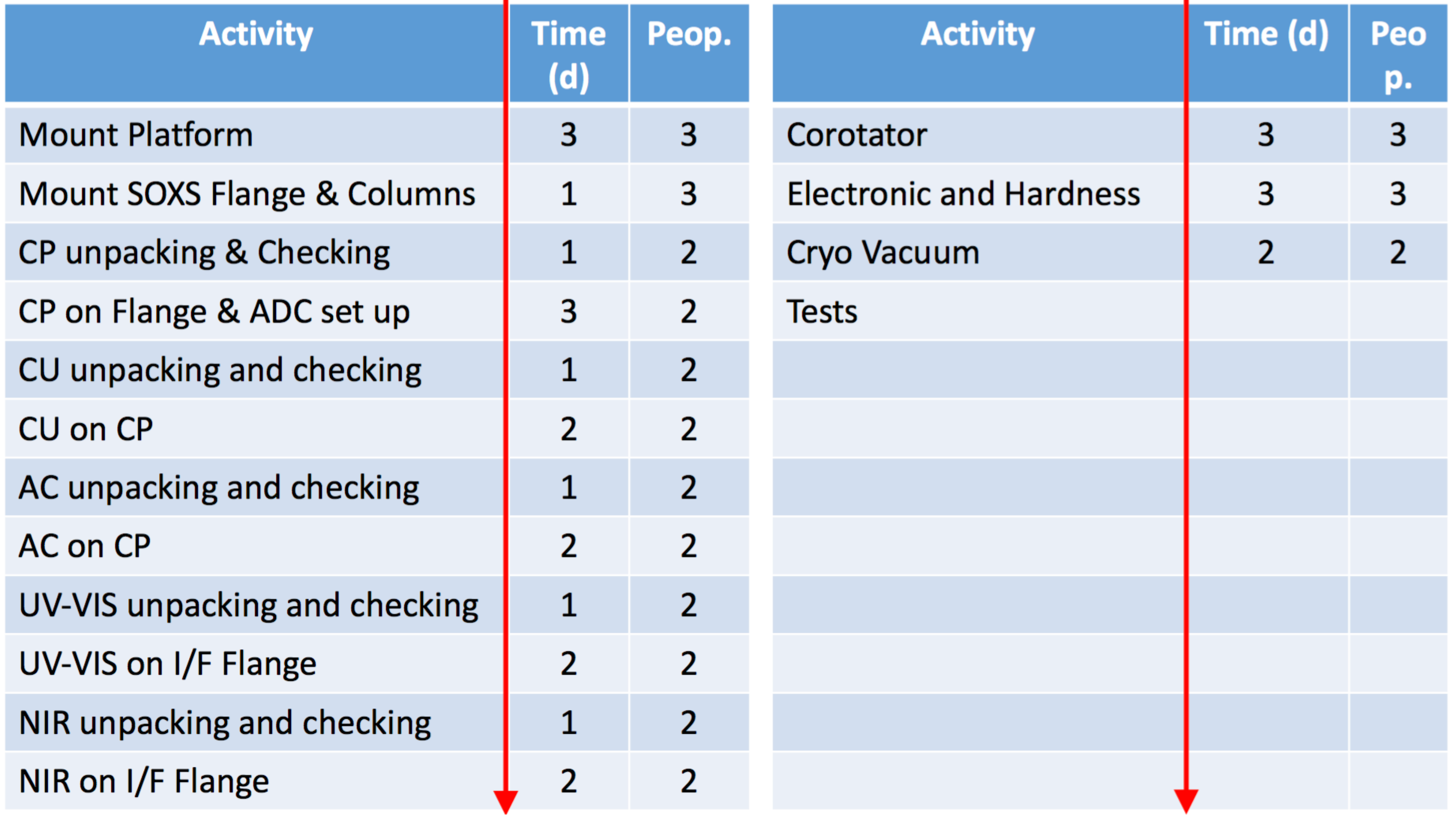}
   \end{tabular}
   \end{center}
   \caption[example] 
   { \label{fig:lasilla} The integration flow of SOXS at la Silla. 
}
   \end{figure} 

The parameter that we cannot control here is the coupling between the I/F flange and the Nasmyth flange, essentially due to different planarity, resulting in a possible misalignment of the instrument w.r.t. the telescope. We will take into account this roto-translation repositioning the flange (shimming) and we will proceed, faster, to the instrument alignment. The first operation at the telescope is the assembly of the new SOXS platform, designed to substitute the SOFI one. The platform components are procured in Milano and sent to La Silla as soon as SOFI is dismount. The whole flux of operation are listed in Figure\ \ref{fig:lasilla}.

\section{Conclusion}
\label{sec:conclusion}  
The SOXS instrument will be used to characterize transient sources. To pursue this aim,  it shall be as simple as possible, rapid in response to an alarm, and efficient. The three scientific arms and the common path of the instrument are built in different places, and all the sub - systems have been gathered in the integration hall of INAF - Padova. This implies a careful management of the integration activities.
In this paper, we presented the flow of the activities from the definition of the technical requirements to the realization of the instrument. Furthermore, as a consequence, we describe the plan for the instrument verification that will constitute most of the activities in preparation for PAE. The integration of SOXS at the telescope (NTT) has also been outlined.

\newpage

\begin{table}[t]
    \centering
    \begin{tabular}{ll}
\multicolumn{2}{c}{\textbf{Abbreviations and Acronyms}} \\   
AD &Applicable Document\\
AI &Action Item\\
CP\_SS &Common Path Selection Slide\\
CRE &Change Request\\
DNO &Discrepancy Note\\
DRD &Document Requirement Description\\
DRH &Data Reduction and Handling\\
DRS &Data Reduction Software\\
ESO &European Southern Observatory\\
FDR &Final Design Review\\
FITS &Flexible Image Transport System\\
FOV &Field Of View\\
FSR &Free Spectral Range\\
HW &HardWare\\
ICD &Interface Control Document\\
ICS &Instrument Control Software\\
KoM &Kick off Meeting\\
LPO &La Silla Paranal Observatory\\
LRU &Line Replaceable Unit\\
MAIT &Manufacture, Assembly, Integration and Test\\
MoU &Memorandum of Understanding\\
NGC &New Generation Controller\\
NTT &New Technology Telescope\\
PAC &Provisional Acceptance Chile\\
PAE &Preliminary Acceptance Europe\\
PDR &Preliminary Design Review\\
PDP &Preliminary Design Phase\\
RAMS &Reliability, Availability, Maintenance and Safety (analysis) RD Reference Document\\
SoW &Statement of Work\\
SOXS &Son Of X-Shooter\\
SW &SoftWare\\
TBC &To Be Confirmed\\
TBD &To Be Defined\\
TCS &Telescope Control System\\
TLR &Top Level Requirements\\
TRS &Technical Requirements Specifications UT Unit Telescope (of the VLT)\\
VLT &Very Large Telescope\\
WP &Work Package\\
    \end{tabular}
    \caption{Abbreviations and acronyms}
    \label{tab:my_label}
\end{table}


\bibliography{report2020} 

\begin{thebibliography}{10}

\bibitem{vernetetal2011}
{Vernet}, J. e.~a., ``{X-shooter, the new wide band intermediate resolution
  spectrograph at the ESO Very Large Telescope},'' {\em A\&A}~{\bf 536},  A105
  (2011).

\bibitem{schipanietal2018spie}
{Schipani}, P., {Campana}, S., {Claudi}, R., {K{\"a}ufl}, H.~U., {Accardo}, M.,
  {Aliverti}, M., {Baruffolo}, A., {Ben Ami}, S., {Biondi}, F., {Brucalassi},
  A., {Capasso}, G., {Cosentino}, R., {D'Alessio}, F., {D'Avanzo}, P.,
  {Hershko}, O., {Gardiol}, D., {Kuncarayacti}, H., {Munari}, M., {Rubin}, A.,
  {Scuderi}, S., {Vitali}, F., {Achr{\'e}n}, J., {Araiza-Duran}, J.~A.,
  {Arcavi}, I., {Bianco}, A., {Cappellaro}, E., {Colapietro}, M., {Della
  Valle}, M., {Diner}, O., {D'Orsi}, S., {Fantinel}, D., {Fynbo}, J.,
  {Gal-Yam}, A., {Genoni}, M., {Hirvonen}, M., {Kotilainen}, J., {Kumar}, T.,
  {Landoni}, M., {Lehti}, J., {Li Causi}, G., {Loreggia}, D., {Marafatto}, L.,
  {Mattila}, S., {Pariani}, G., {Pignata}, G., {Rappaport}, M., {Ricci}, D.,
  {Riva}, M., {Salasnich}, B., {Zanmar Sanchez}, R., {Smartt}, S., and
  {Turatto}, M., ``{SOXS: a wide band spectrograph to follow up transients},''
  in [{\em Ground-based and Airborne Instrumentation for Astronomy
  VII}{\nolinebreak\hspace{0.1em}]},  {Evans}, C.~J., {Simard}, L., and
  {Takami}, H., eds., {\em Society of Photo-Optical Instrumentation Engineers
  (SPIE) Conference Series} {\bf 10702},  107020F (July 2018).

\bibitem{schipanietal2016spie}
{Schipani}, P., {Claudi}, R., {Campana}, S., {Baruffolo}, A., {Basa}, S.,
  {Basso}, S., {Cappellaro}, E., {Cascone}, E., {Cosentino}, R., {D'Alessio},
  F., {De Caprio}, V., {Della Valle}, M., {de Ugarte Postigo}, A., {D'Orsi},
  S., {Franzen}, R., {Fynbo}, J., {Gal-Yam}, A., {Gardiol}, D., {Giro}, E.,
  {Hamuy}, M., {Iuzzolino}, M., {Loreggia}, D., {Mattila}, S., {Munari}, M.,
  {Pignata}, G., {Riva}, M., {Savarese}, S., {Schmidt}, B., {Scuderi}, S.,
  {Smartt}, S., and {Vitali}, F., ``{The new SOXS instrument for the ESO
  NTT},'' in [{\em Ground-based and Airborne Instrumentation for Astronomy
  VI}{\nolinebreak\hspace{0.1em}]},  {Evans}, C.~J., {Simard}, L., and
  {Takami}, H., eds., {\em Society of Photo-Optical Instrumentation Engineers
  (SPIE) Conference Series} {\bf 9908},  990841 (Aug. 2016).

\bibitem{claudietal2018frap}
{Claudi}, R., {Campana}, S., {Schipani}, P., {Aliverti}, M., {Baruffolo}, A.,
  {Ben-Ami}, S., {Biondi}, F., {Brucalassi}, A., {Capasso}, G., {Cosentino},
  R., {D'Alessio}, F., {D'Avanzo}, P., {Hershko}, O., {Kuncarayakti}, H.,
  {Munari}, M., {Rubin}, A., {Scuderi}, S., {Vitali}, F., {Achr{\'e}n}, J.,
  {Arcavi}, I., {Duran}, J.~A.~A., {Bianco}, A., {Cappellaro}, E.,
  {Colapietro}, M., {Diner}, O., {Valle}, M.~D., {D'Orsi}, S., {Fantinel}, D.,
  {Fynbom}, J., {Gal-Yam}, A., {Genoni}, M., {Hirvonen}, M., {Kotilainen}, J.,
  {Kumar}, T., {Landoni}, M., {Lehti}, J., {Marafatto}, L., {Causi}, G.~L.,
  {Mattila}, S., {Pariani}, G., {Pignata}, G., {Rappaport}, M., {Riva}, M.,
  {Ricci}, D., {Salasnich}, B., {Sanchez}, R., {Smartt}, S., {Turatto}, M.,
  {K{\"a}ufle}, H.~U., and {Accardo}, M., ``{Son of X-Shooter: a multi-band
  instrument for a multi-band universe},'' in [{\em Frontier Research in
  Astrophysics - III}{\nolinebreak\hspace{0.1em}]},   78 (June 2018).

\bibitem{schipanietal2020spie}
{Schipani}, P. e.~a., ``{Development status of the SOXS spectrograph for the
  ESO-NTT Telescope},'' in [{\em {Telescope and Astronomical Instrumentation
  2020}}{\nolinebreak\hspace{0.1em}]},  {\em Proc. SPIE 11447} (2020).

\bibitem{rubinetal2018spie}
{Rubin}, A., {Ben-Ami}, S., {Hershko}, O., {Rappaport}, M., {Diner}, O.,
  {Gal-Yam}, A., {Campana}, S., {Claudi}, R., {Schipani}, P., {Aliverti}, M.,
  {Baruffolo}, A., {Biondi}, F., {Brucalassi}, A., {Capasso}, G., {Cosentino},
  R., {D'Alessio}, F., {D'Avanzo}, P., {Kuncarayakti}, H., {Munari}, M.,
  {Scuderi}, S., {Vitali}, F., {Achr{\'e}n}, J., {Araiza-Duran}, J.~A.,
  {Arcavi}, I., {Bianco}, A., {Cappellaro}, E., {Colapietro}, M., {Della
  Valle}, M., {D'Orsi}, S., {Fantinel}, D., {Fynbo}, J., {Genoni}, M.,
  {Hirvonen}, M., {Kotilainen}, J., {Kumar}, T., {Landoni}, M., {Lehti}, J.,
  {Li Causi}, G., {Marafatto}, L., {Mattila}, S., {Pariani}, G., {Pignata}, G.,
  {Ricci}, D., {Riva}, M., {Salasnich}, B., {Zanmar Sanchez}, R., {Smartt}, S.,
  and {Turatto}, M., ``{MITS: the multi-imaging transient spectrograph for
  SOXS},'' in [{\em Ground-based and Airborne Instrumentation for Astronomy
  VII}{\nolinebreak\hspace{0.1em}]},  {Evans}, C.~J., {Simard}, L., and
  {Takami}, H., eds., {\em Society of Photo-Optical Instrumentation Engineers
  (SPIE) Conference Series} {\bf 10702},  107022Z (July 2018).

\bibitem{rubinetal2020spie}
{Rubin}, A. e.~a., ``{Progress on the UV-VIS arm of SOXS},'' in [{\em
  {Telescope and Astronomical Instrumentation
  2020}}{\nolinebreak\hspace{0.1em}]},  {\em Proc. SPIE 11447} (2020).

\bibitem{cosentinoetal2018spie}
{Cosentino}, R., {Aliverti}, M., {Scuderi}, S., {Campana}, S., {Claudi}, R.,
  {Schipani}, P., {Baruffolo}, A., {Ben-Ami}, S., {Mehrgan}, L.~H., {Ives}, D.,
  {Biondi}, F., {Brucalassi}, A., {Capasso}, G., {D'Alessio}, F., {D'Avanzo},
  P., {Diner}, O., {Kuncarayakti}, H., {Munari}, M., {Rubin}, A., {Vitali}, F.,
  {Achr{\'e}n}, J., {Araiza-Dur{\'a}n}, J.~A., {Arcavi}, I., {Bianco}, A.,
  {Cappellaro}, E., {Colapietro}, M., {Della Valle}, M., {D'Orsi}, S.,
  {Fantinel}, D., {Fynbo}, J., {Gal-Yam}, A., {Genoni}, M., {Hirvonen}, M.,
  {Kotilainen}, J., {Kumar}, T., {Landoni}, M., {Lehti}, J., {Li Causi}, G.,
  {Marafatto}, L., {Mattila}, S., {Pariani}, G., {Pignata}, G., {Rappaport},
  M., {Ricci}, D., {Riva}, M., {Salasnich}, B., {Zanmar Sanchez}, R., {Smartt},
  S., and {Turatto}, M., ``{The VIS detector system of SOXS},'' in [{\em
  Ground-based and Airborne Instrumentation for Astronomy
  VII}{\nolinebreak\hspace{0.1em}]},  {Evans}, C.~J., {Simard}, L., and
  {Takami}, H., eds., {\em Society of Photo-Optical Instrumentation Engineers
  (SPIE) Conference Series} {\bf 10702},  107022J (July 2018).

\bibitem{cosentinoetal2020spie}
{Cosentino}, R. e.~a., ``{Development status of the UV-VIS detector system of
  SOXS for the ESO NTT Telescope},'' in [{\em {Telescope and Astronomical
  Instrumentation 2020}}{\nolinebreak\hspace{0.1em}]},  {\em Proc. SPIE 11447}
  (2020).

\bibitem{vitalietal2018spie}
{Vitali}, F., {Aliverti}, M., {Capasso}, G., {D'Alessio}, F., {Munari}, M.,
  {Riva}, M., {Scuderi}, S., {Zanmar Sanchez}, R., {Campana}, S., {Schipani},
  P., {Claudi}, R., {Baruffolo}, A., {Ben-Ami}, S., {Biondi}, F., {Brucalassi},
  A., {Cosentino}, R., {Ricci}, D., {D'Avanzo}, P., {Diner}, O.,
  {Kuncarayakti}, H., {Rubin}, A., {Achr{\'e}n}, J., {Araiza-Duran}, J.~A.,
  {Arcavi}, I., {Bianco}, A., {Cappellaro}, E., {Colapietro}, M., {Della
  Valle}, M., {D'Orsi}, S., {Fantinel}, D., {Fynbo}, J., {Gal-Yam}, A.,
  {Genoni}, M., {Hirvonen}, M., {Kotilainen}, J., {Kumar}, T., {Landoni}, M.,
  {Lehti}, J., {Li Causi}, G., {Marafatto}, L., {Mattila}, S., {Pariani}, G.,
  {Pignata}, G., {Rappaport}, M., {Salasnich}, B., {Smartt}, S., and {Turatto},
  M., ``{The NIR spectrograph for the new SOXS instrument at the NTT},'' in
  [{\em Ground-based and Airborne Instrumentation for Astronomy
  VII}{\nolinebreak\hspace{0.1em}]},  {Evans}, C.~J., {Simard}, L., and
  {Takami}, H., eds., {\em Society of Photo-Optical Instrumentation Engineers
  (SPIE) Conference Series} {\bf 10702},  1070228 (July 2018).

\bibitem{vitalietal2020spie}
{Vitali}, F. e.~a., ``{The development status of the NIR spectrograph for the
  new SOXS instrument at the NTT},'' in [{\em {Telescope and Astronomical
  Instrumentation 2020}}{\nolinebreak\hspace{0.1em}]},  {\em Proc. SPIE 11447,
  AS105-287} (2018).

\bibitem{brucalassietal2018spie}
{Brucalassi}, A., {Araiza-Dur{\'a}n}, J.~A., {Pignata}, G., {Campana}, S.,
  {Claudi}, R., {Schipani}, P., {Aliverti}, M., {Baruffolo}, A., {Ben-Ami}, S.,
  {Biondi}, F., {Capasso}, G., {Cosentino}, R., {D'Alessio}, F., {D'Avanzo},
  P., {Gardiol}, D., {Hershko}, O., {Kuncarayakti}, H., {Munari}, M., {Ricci},
  D., {Riva}, M., {Rubin}, A., {Zanmar Sanchez}, R., {Scuderi}, S., {Vitali},
  F., {Achr{\'e}n}, J., {Arcavi}, I., {Bianco}, A., {Cappellaro}, E.,
  {Colapietro}, M., {Della Valle}, M., {Diner}, O., {D'Orsi}, S., {Fantinel},
  D., {Fynbo}, J., {Gal-Yam}, A., {Genoni}, M., {Hirvonen}, M., {Kotilainen},
  J., {Kumar}, T., {Landoni}, M., {Lehti}, J., {Li Causi}, G., {Loreggia}, D.,
  {Marafatto}, L., {Mattila}, S., {Pariani}, G., {Rappaport}, M., {Salasnich},
  B., {Smartt}, S., and {Turatto}, M., ``{The acquisition camera system for
  SOXS at NTT},'' in [{\em Ground-based and Airborne Instrumentation for
  Astronomy VII}{\nolinebreak\hspace{0.1em}]},  {Evans}, C.~J., {Simard}, L.,
  and {Takami}, H., eds., {\em Society of Photo-Optical Instrumentation
  Engineers (SPIE) Conference Series} {\bf 10702},  107022M (July 2018).

\bibitem{brucalassietal2020spie}
{Brucalassi}, A. e.~a., ``{Final design and development status of the
  Acquisition and Guiding system for SOXS},'' in [{\em {Telescope and
  Astronomical Instrumentation 2020}}{\nolinebreak\hspace{0.1em}]},  {\em Proc.
  SPIE 11447} (2020).

\bibitem{kuncarayaktietal2020spie}
{Kuncarayakti}, H. e.~a., ``{Design and development of the SOXS calibration
  unit},'' in [{\em {Telescope and Astronomical Instrumentation
  2020}}{\nolinebreak\hspace{0.1em}]},  {\em Proc. SPIE 11447} (2020).

\bibitem{claudietal2018spie}
{Claudi}, R., {Aliverti}, M., {Biondi}, F., {Munari}, M., {Zanmar Sanchez}, R.,
  {Campana}, S., {Schipani}, P., {Baruffolo}, A., {Ben-Ami}, S., {Brucalassi},
  A., {Capasso}, G., {Cosentino}, R., {D'Alessio}, F., {D'Avanzo}, P.,
  {Hershko}, O., {Kuncarayakti}, H., {Rubin}, A., {Scuderi}, S., {Vitali}, F.,
  {Achr{\'e}n}, J., {Araiza-Dur{\'a}n}, J.~A., {Arcavi}, I., {Bianco}, A.,
  {Cappellaro}, E., {Colapietro}, M., {Della Valle}, M., {Diner}, O., {D'Orsi},
  S., {Fantinel}, D., {Fynbo}, J., {Gal-Yam}, A., {Genoni}, M., {Hirvonen}, M.,
  {Kotilainen}, J., {Kumar}, T., {Landoni}, M., {Lehti}, J., {Li Causi}, G.,
  {Marafatto}, L., {Mattila}, S., {Pariani}, G., {Pignata}, G., {Rappaport},
  M., {Ricci}, D., {Riva}, M., {Salasnich}, B., {Smartt}, S., and {Turatto},
  M., ``{The common path of SOXS (Son of X-Shooter)},'' in [{\em Ground-based
  and Airborne Instrumentation for Astronomy VII}{\nolinebreak\hspace{0.1em}]},
   {Evans}, C.~J., {Simard}, L., and {Takami}, H., eds., {\em Society of
  Photo-Optical Instrumentation Engineers (SPIE) Conference Series} {\bf
  10702},  107023T (July 2018).

\bibitem{biondietal2018spie}
{Biondi}, F., {Claudi}, R., {Marafatto}, L., {Farinato}, J., {Magrin}, D.,
  {Ragazzoni}, R., {Campana}, S., {Schipani}, P., {Aliverti}, M., {Baruffolo},
  A., {Ben-Ami}, S., {Brucalassi}, A., {Capasso}, G., {Cosentino}, R.,
  {D'Alessio}, F., {D'Avanzo}, P., {Hershko}, O., {Kuncarayakti}, H., {Munari},
  M., {Rubin}, A., {Scuderi}, S., {Vitali}, F., {Achr{\'e}n}, J.,
  {Araiza-Dur{\'a}n}, J.~A., {Arcavi}, I., {Bianco}, A., {Cappellaro}, E.,
  {Colapietro}, M., {Della Valle}, M., {Diner}, O., {D'Orsi}, S., {Fantinel},
  D., {Fynbo}, J., {Gal-Yam}, A., {Genoni}, M., {Hirvonen}, M., {Kotilainen},
  J., {Kumar}, T., {Landoni}, M., {Lehti}, J., {Li Causi}, G., {Mattila}, S.,
  {Pariani}, G., {Pignata}, G., {Rappaport}, M., {Riva}, M., {Ricci}, D.,
  {Salasnich}, B., {Zanmar Sanchez}, R., {Smartt}, S., and {Turatto}, M.,
  ``{The assembly integration and test activities for the new SOXS instrument
  at NTT},'' in [{\em Ground-based and Airborne Instrumentation for Astronomy
  VII}{\nolinebreak\hspace{0.1em}]},  {Evans}, C.~J., {Simard}, L., and
  {Takami}, H., eds., {\em Society of Photo-Optical Instrumentation Engineers
  (SPIE) Conference Series} {\bf 10702},  107023D (July 2018).

\bibitem{biondietal2020spie}
{Biondi}, F. e.~a., ``{The AIV strategy of the Common Path of Son of
  X-Shooter},'' in [{\em {Telescope and Astronomical Instrumentation
  2020}}{\nolinebreak\hspace{0.1em}]},  {\em Proc. SPIE 11447} (2020).

\bibitem{claudietal2019frapconfe78}
{Claudi}, R., {Campana}, S., {Schipani}, P., {Aliverti}, M., {Baruffolo}, A.,
  {Ben-Ami}, S., {Biondi}, F., {Brucalassi}, A., {Capasso}, G., {Cosentino},
  R., {D'Alessio}, F., {D'Avanzo}, P., {Hershko}, O., {Kuncarayakti}, H.,
  {Munari}, M., {Rubin}, A., {Scuderi}, S., {Vitali}, F., {Achr{\'e}n}, J.,
  {Arcavi}, I., {Duran}, J.~A.~A., {Bianco}, A., {Cappellaro}, E.,
  {Colapietro}, M., {Diner}, O., {Valle}, M.~D., {D'Orsi}, S., {Fantinel}, D.,
  {Fynbom}, J., {Gal-Yam}, A., {Genoni}, M., {Hirvonen}, M., {Kotilainen}, J.,
  {Kumar}, T., {Landoni}, M., {Lehti}, J., {Marafatto}, L., {Causi}, G.~L.,
  {Mattila}, S., {Pariani}, G., {Pignata}, G., {Rappaport}, M., {Riva}, M.,
  {Ricci}, D., {Salasnich}, B., {Sanchez}, R., {Smartt}, S., {Turatto}, M.,
  {K{\"a}ufle}, H.~U., and {Accardo}, M., ``{Son of X-Shooter: a multi-band
  instrument for a multi-band universe},'' in [{\em Frontier Research in
  Astrophysics - III. 28 May - 2 June 2018. Mondello
  (Palermo}{\nolinebreak\hspace{0.1em}]},   78 (Nov. 2019).

\end{thebibliography}
\bibliographystyle{spiebib} 

\end{document}